\documentclass[twocolumn]{aastex631}

\usepackage{soul}
\usepackage{natbib}
\usepackage{multirow}
\usepackage{graphicx}
\usepackage{tabularx}
\usepackage{verbatim}
\usepackage{color}
\usepackage{ulem}
\usepackage{subfigure}

\newcommand\soutm{\bgroup\markoverwith
{\textcolor{black}{\rule[0.5ex]{2pt}{0.8pt}}}\ULon}

\submitjournal{ApJ}

\shorttitle{The Galaxy Replacement Technique}
\shortauthors{Chun et al.}

\graphicspath{{./}{figures/}}

\begin{document}

\title{The Galaxy Replacement Technique (GRT): a New Approach to Study Tidal Stripping and Formation of Intracluster Light in a Cosmological Context}

\correspondingauthor{Kyungwon Chun}
\email{kwchun@kasi.re.kr}

\author{Kyungwon Chun}
\affil{Korea Astronomy and Space Science Institute (KASI), 776 Daedeokdae-ro, Yuseong-gu, Daejeon 34055, Korea}
\author{Jihye Shin}
\affil{Korea Astronomy and Space Science Institute (KASI), 776 Daedeokdae-ro, Yuseong-gu, Daejeon 34055, Korea}
\author{Rory Smith}
\affil{Korea Astronomy and Space Science Institute (KASI), 776 Daedeokdae-ro, Yuseong-gu, Daejeon 34055, Korea}
\affil{University of Science and Technology (UST), Gajeong-ro, Daejeon 34113, Korea}
\author{Jongwan Ko}
\affil{Korea Astronomy and Space Science Institute (KASI), 776 Daedeokdae-ro, Yuseong-gu, Daejeon 34055, Korea}
\affil{University of Science and Technology (UST), Gajeong-ro, Daejeon 34113, Korea}
\author{Jaewon Yoo}
\affil{Korea Astronomy and Space Science Institute (KASI), 776 Daedeokdae-ro, Yuseong-gu, Daejeon 34055, Korea}
\affil{University of Science and Technology (UST), Gajeong-ro, Daejeon 34113, Korea}

\begin{abstract}
We introduce the Galaxy Replacement Technique (GRT) that allows us to model tidal stripping of galaxies with very high-mass (m$_{\rm{star}}=5.4\times10^4$~M$_\odot$/h) and high-spatial resolution (10 pc/h), in a fully cosmological context, using an efficient and fast technique. The technique works by replacing multiple low-resolution DM halos in the base cosmological simulation with high-resolution models, including a DM halo and stellar disk. We apply the method to follow the hierarchical build-up of a cluster since redshift $\sim8$ to now, through the hierarchical accretion of galaxies, individually or in substructures such as galaxy groups. We find we can successfully reproduce the observed total stellar masses of observed clusters since redshift $\sim$1. The high resolution allows us to accurately resolve the tidal stripping process and well describe the formation of ultra-low surface brightness features in the cluster ($\mu_{V}<32$ mag arcsec$^{-2}$) such as the intracluster light (ICL), shells and tidal streams. We measure the evolution of the fraction of light in the ICL and brightest cluster galaxy (BCG) using several different methods. While their broad response to the cluster mass growth history is similar, the methods show systematic differences, meaning we must be careful when comparing studies that use distinct methods. The GRT represents a powerful new tool for studying tidal effects on galaxies and exploring the formation channels of the ICL in a fully cosmological context and with large samples of simulated groups and clusters.

\end{abstract}


\section{Introduction}
In the current concordance cosmology of Lambda cold dark matter ($\Lambda$CDM), galaxy clusters are the largest gravitationally-bound structures, hierarchically built up by merging of smaller structures. 
Some of the accreted structures can survive until today, such as satellite galaxies, but are tidally disrupted and spread throughout the galaxy cluster's potential well.
Among these disrupted components, the luminous components are defined as the intracluster light (ICL).

Since the ICL was first observed in the Coma cluster by \citet{zwicky1951}, the ICL has been observed in many nearby clusters \citep{mihos2005,janowiecki2010,rudick2010,jimenez-teja2019} and even in clusters beyond $z=1$ \citep{adami2013,ko2018}.
Recent observational studies have shown that the ICL is a ubiquitous component, but the amount of the ICL varies between clusters and with redshift, from 2\% to 40\% of the total stellar mass in the clusters \citep{burke2015,morishita2017,jimenez-teja2018,montes2018,furnell2021}.
The fact that the ICL occupies a significant fraction of the total stellar mass in the clusters indicates that studying the ICL can give us insight into the formation and evolution of the clusters.

Previous studies have suggested several channels for the formation of the ICL, such as in-situ star formation \citep{puchwein2010}, tidal disruption of the dwarf galaxies \citep{janowiecki2010}, tidal stripping of satellite galaxies \citep{rudick2009,puchwein2010,contini2018,montes2018}, major mergers \citep{murante2007,contini2018}, and pre-processing in groups \citep{rudick2006}.
These formation channels contribute to the formation of the ICL in all clusters, but their relative importance may differ depending on the dynamical state and properties of the clusters.
However, the most dominant mechanism is not yet clear.
Furthermore, studies have generally not found consistent results about the relation between the ICL and the properties of the clusters.
\citet{burke2015} and \citet{furnell2021} showed that the amount of the ICL increases with time until $z\sim0.1$. 
In contrast, \citet{montes2018} showed that the ICL fraction is not significantly increasing between $0.3 < z < 0.6$  using six massive clusters, and \citet{morishita2017} also found a similar trend.
Inconsistent results between studies have also been revealed in the relation between the fraction of ICL, the cluster dynamical state \citep{montes2018,jimenez-teja2018,canas2020}, and the cluster mass \citep{purcell2007,contini2014,cui2014,asensio2020}.

The disagreements could result from the unique properties of all the clusters in the previous studies but might be due to the different detection methods and definitions of the ICL. 
Indeed, \citet{montes2018} showed that applying two distinct methods to define the ICL resulted in significantly different amounts and evolution of the ICL in six massive observed clusters.
Using controlled N-body simulations, \citet{rudick2011} found that the ICL fraction of clusters can change by up to a factor of three due to differing definitions.
Thus, a statistical ICL analysis requires large cluster samples, combined with a matching definition of the ICL.

Previous numerical studies for the ICL usually have been conducted using either cosmological hydrodynamic simulations or a semi-analytic model based on cosmological N-body simulations \citep{conroy2007,murante2007,dolag2010,contini2014,contini2018,cui2014,pillepich2018,tang2018,canas2020}.
Studies using cosmological hydrodynamic simulations trace the formation and evolution of the stellar component of the ICL self-consistently, with stars forming from the gas component whose hydrodynamics is modelled.
However, these treatments of the baryonic component are highly computationally expensive. As a result, the number of clusters considered or the mass or spatial resolution of the simulations is necessarily limited. 

Semi-analytical models do not include the baryonic component self-consistently, instead of using semi-analytical recipes for how the baryons should be traced onto the dark matter halos. This makes them relatively fast compared to the cosmological hydrodynamic simulations, meaning much larger cosmological volumes and larger samples of clusters can be considered. However, the predictions about the properties of the ICL are limited. For example, while the amount of ICL in a cluster may be predicted, little is known about its spatial distribution.
Some previous studies trace the stellar distribution using the particle tagging method with semi-analytic models from Milky Way to cluster scales \citep{bullock2005,cooper2010,cooper2015}, however, the mass and spatial resolution are limited to the original DM-only simulation from which the tagged particles are selected.

Unlike these standard methods to study the ICL, some studies have investigated the low-surface brightness tidal features and ICL that contribute to the diffuse light using controlled simulations, either in a cosmological or non-cosmological context \citep{rudick2006,rudick2009,rudick2011,laporte2013,ji2014,yozin2015,mancillas2019}.
These simulations show the time and spatial evolution of the structures in the clusters with high resolution.
Moreover, these simulation methods enable us to study the ICL and low-surface brightness tidal features systematically because their controlled nature means that the user can choose the parameters of the cluster considered. However, these non-cosmological simulations were generally unable to fully take into account their cosmological context.
Meanwhile, those that were set in a cosmological context were limited by their design choices such as the model of galaxies that contribute to the formation of ICL and the starting point of the re-simulation. We discuss these limitations in more detail in Section \ref{sec:rudick}.

In this work, we introduce an alternative simulation technique referred to as the `Galaxy Replacement Technique' (GRT). 
The GRT focuses on following the build-up of a cluster with a multi-resolution cosmological N-body re-simulation, derived from the full merger tree of a low-resolution DM-only cosmological simulation of a cluster.
As this technique utilizes a re-simulation approach, we can choose and control the properties and time evolution of the cluster while still fully considering their cosmological context.
This means that we can target various clusters with, for example, particular merger histories or environments, depending on the focus of our study.
Moreover, the GRT traces the spatial distribution and evolution of the cluster and its sub-structures without the inclusion of computationally expensive baryonic physics.
This inexpensive calculation enables us to study large statistical samples of clusters with both high mass and high spatial resolution, which is critical for providing predictions on very low surface brightness features. The high spatial resolution allows us to resolve the tidal stripping process accurately, and the high mass resolution allows us to model diffuse features with sufficient star particles. Thus, the GRT is ideal for a statistical study of ICL formation in a cosmological context.

This paper is organized as follows.
In Section \ref{sec:GRT}, we introduce the galaxy replacement technique, which is developed to trace the evolution of the ICL and describe the properties of a single cluster in a multi-resolution re-simulation.
Section \ref{sec:test} is devoted to conducting tests of the GRT to show that it is well suited for studying the ICL.
In Section \ref{sec:ICL}, we describe various detection methods for quantifying the ICL and the BCG and put all this together for a single cluster to see the time evolution of the ICL and BCG fraction.
Lastly, we discuss the simulation technique and future works in Section \ref{sec:discussion} and summarize in Section \ref{sec:summary}.

\section{The Galaxy Replacement Technique}
\label{sec:GRT}

\begin{figure*}
\centering
\includegraphics[width=0.9\textwidth]{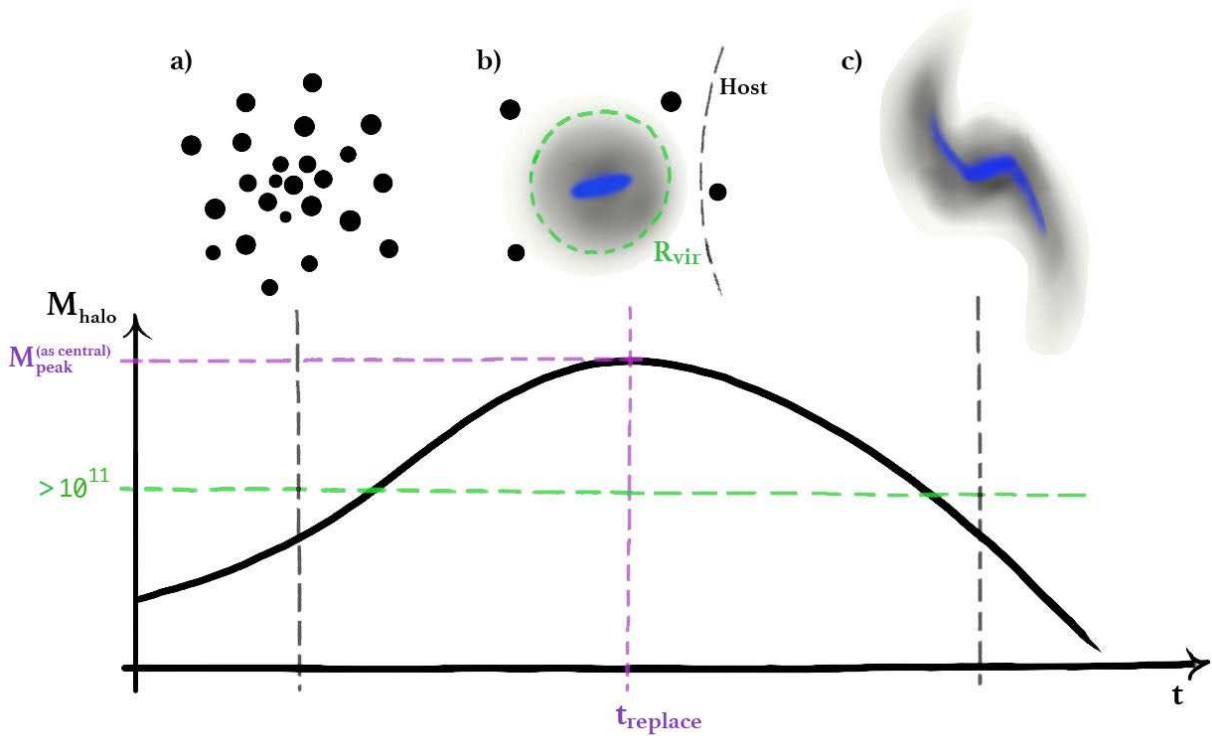}
\caption{Cartoon schematic of the time evolution of a halo that reaches M$_{peak}$ at t$_{replace}$. a) The halo is composed of the low-resolution DM particles (the black filled circles) until t$_{replace}$. b) The particles in the virial radius of the halo (the green dashed circle) are replaced by a high-resolution DM (the gray scale) and stellar disk (the blue scale), and c) then fall into the host halo where they suffer tidal deformation and stripping.}
\label{fig:GRT}
\end{figure*}

The GRT enables us to trace the spatial distribution and evolution of the ICL without the computationally expensive baryonic physics. The outline of the GRT for the ICL study can be simply described as following steps. (1) A low-resolution DM-only cosmological simulation is performed. (2) A galaxy cluster for re-simulation is selected, and a merger tree of all progenitors is built. (3) Using the merger trees, a DM halo of progenitors that satisfy the GRT criteria are substituted by high-resolution DM halo and galaxy. The stellar mass of this galaxy is decided based on the halo abundance matching. (4) A multi-resolution re-simulation is performed.

A low-resolution DM-only cosmological simulation of (120~Mpc$/h$)$^3$ uniform box is performed with 512$^3$ particles from $z=200$ to 0 using the cosmological simulation code, Gadget-3 \citep{springel2005}.
Hereafter, we call this low-resolution DM-only simulation the `base simulation'.
The position and velocity of DM particles are stored in 170 discrete snapshots.
The time interval between two snapshots is increasing from 16~Myr to 100~Myr until $z=1$ and it is fixed as 100~Myr after then.
In the base simulation, the particle mass is $\sim$ 10$^9$~M$_{\odot}/h$, and the gravitational softening length is fixed at 2.3 kpc/h on a comoving scale.
For this simulation, we use the post-Planck cosmological model of $\Omega_{m} = 0.3,~\Omega_{\Lambda} = 0.7,~\Omega_{b} = 0.047$, and $h = 0.684$.
Initial condition for particle positions and velocities is generated by a MUSIC package\footnote{https://bitbucket.org/ohahn/music/} \citep{hahn2011}.
The halo and subhalo structures are identified with the modified six-dimensional phase-space halo finder ROCKSTAR\footnote{https://bitbucket.org/pbehroozi/rockstar-galaxies} \citep{behroozi2013a}, which can consider multiple types of particles simultaneously, such as dark matter, gas, and stars. 
The lower mass limit of a halo is $\sim$ 2$\times$10$^{10}$~M$_{\odot}/h$ (N$_\textrm{DM}=20$).
The merger tree of these halos is built using Consistent Trees \citep{behroozi2013b}.
The halos are linked with their progenitors in all snapshots to form the merger tree. 
From the merger tree, we identify the mass growth history of all halos that will later constitute the cluster and its subhalos, regardless of whether the halos survive until $z=0$ or not. 

Because the box size of the base simulation is large enough, more than 40 cluster mass halos (M$_{vir}>$10$^{14}$~M$_{\odot}/h$ at z=0) naturally form within it. 
To study the ICL, we select a single target cluster of M$_{vir}\sim$2$\times$10$^{14}$~M$_{\odot}/h$ at $z=0$ from the base simulation, that we will use as an example in Section \ref{sec:fraction}. We refer to this target cluster as the `GRT cluster'.

For the GRT cluster, we only replace low-resolution halos with M$_{peak}>$10$^{11}$~M$_{\odot}/h$, where M$_{peak}$ is the maximum mass of a DM halo while it is still a central halo (Figure \ref{fig:GRT}). Halos with M$_{peak}<$10$^{11}$~M$_{\odot}/h$ consist of less than 100 low-resolution DM particles, and we find they contribute less than 2$~\%$ of the total stellar mass in the clusters at $z=0$, thus we do not replace them to reduce the overall computational cost.

The halos with M$_{peak}>$10$^{11}$~M$_{\odot}/h$ remain as the low-resolution DM halos until their mass reach M$_{peak}$ but then are replaced by a high-resolution model that consists of a high-resolution DM halo and a stellar disk composed of high-resolution star particles (Figure \ref{fig:GRT}a, b). Hereafter, we refer to the low-resolution DM halo as the `low-res. halo' and the substituted high-resolution DM halo and luminous galaxy as the `high-res. galaxy'. In this replacement process, only the particles inside the virial radius of the low-res. halo will be replaced with high-resolution particles. As particles outside the high-res. galaxy remain as low-resolution particles, we refer to this as a multi-resolution re-simulation (Figure \ref{fig:GRT}b). We also tested replacing all low-resolution particles out to two virial radii but found it made a negligible difference to our results.
In principle, the high-res. galaxy could be contaminated by the inflow of the low-resolution particles, which might result in artificial heating effects. However, by replacing the galaxies when their halos reach their peak mass, we significantly reduce the amount of low-resolution particles bound to the high-res. galaxy. We run tests on artificial heating of the stellar disk by low-resolution particles in Section \ref{sec:heating}, and verify that our approach is successful at limiting the artificial heating, and that this issue is not significant for our ICL studies.

For each high-res. galaxy, the initial position, and velocity of the high-resolution particles in the galaxy's own frame of reference are generated by DICE\footnote{https://bitbucket.org/vperret/dice} \citep{perret2014}.
DICE generates the initial conditions of idealized galaxies in dynamical equilibrium with multiple components, including a bulge, thick/thin disk, and halo. We can parameterize the various properties (the mass of particles, density model, shape, etc.) of each component.
We assume an NFW density profile for the high-resolution DM halos.
Many previous studies have shown that low-mass halos ($<$10$^{11}$~M$_{\odot}$) have a flatter core due to the stellar feedback \citep{dicintio2014,dutton2016,tollet2016,freundlich2020}. This decrease of central density will have implications for the tidal disruption \citep{penarrubia2008,smith2016}, but we do not expect that this is very significant for our studies of the ICL formation because all of our high resolution model galaxies are more massive than 10$^{11}$~M$_{\odot}$.
The halo spin parameter and concentration are matched to the ROCKSTAR properties of the replaced low-res. halo.
We tested the galaxy models generated by DICE in isolation using an idealized simulation and find that the initial conditions are very close to equilibrium and evolve in a highly stable manner.
Because the low-res. halo is replaced with a high-res. galaxy before the halo interacts with another massive halo, we assume the halo has only one luminous galaxy. However, host galaxies with satellites can naturally form later via the normal process of hierarchical accretion in a cosmological context. 

To substantially reduce the numerical cost, we do not consider star formation. Instead, by inserting fully formed disks at the moment the halo peaks in mass, we essentially capture all of the star formation up until the moment of infall. This has advantages because we can ensure the amount of stars produce at the time of infall is an exact match to the observed stellar halo-mass relation at all masses and redshifts. Indeed, this is a great challenge for fully hydrodynamical cosmological simulations to meet that is under our control in the GRT. We apply the stellar mass-halo mass relation as a function of redshift from \cite{behroozi2013c}. We examine this specific choice of abundance matching in Section \ref{sec:am}. After the peak mass is reached, which generally coincides with the galaxies entering a group or cluster mass host, there is no further stellar mass growth. However, most studies have shown that star formation is quickly and significantly suppressed or halted in dense environments, and so the fraction of stars that would be produced after entering these environments can be expected to be only a small fraction of the total \citep{koopermann2004,oman2016,smith2016,smith2019,lotz2019,rhee2020,mostoghiu2021}.

We assume that the inserted galaxy is a bulge-less exponential disk galaxy. This is reasonable as most low-mass disk galaxies with M$_{star} < $10$^{10}$~M$_{\odot}$ have this morphology \citep{dutton2009}. However, we note that bulges and spheroids, and other dispersion-supported structures will naturally evolve in our simulation as galaxies merge and tidally interact.
Also, as some galaxies are inserted outside of the cluster (for example, after entering groups), we can track the formation of diffuse light in environments outside of the main cluster that are then later brought into the cluster during its hierarchical build-up (Figure \ref{fig:GRT}c).

\begin{figure*}
\centering
\includegraphics[width=0.9\textwidth]{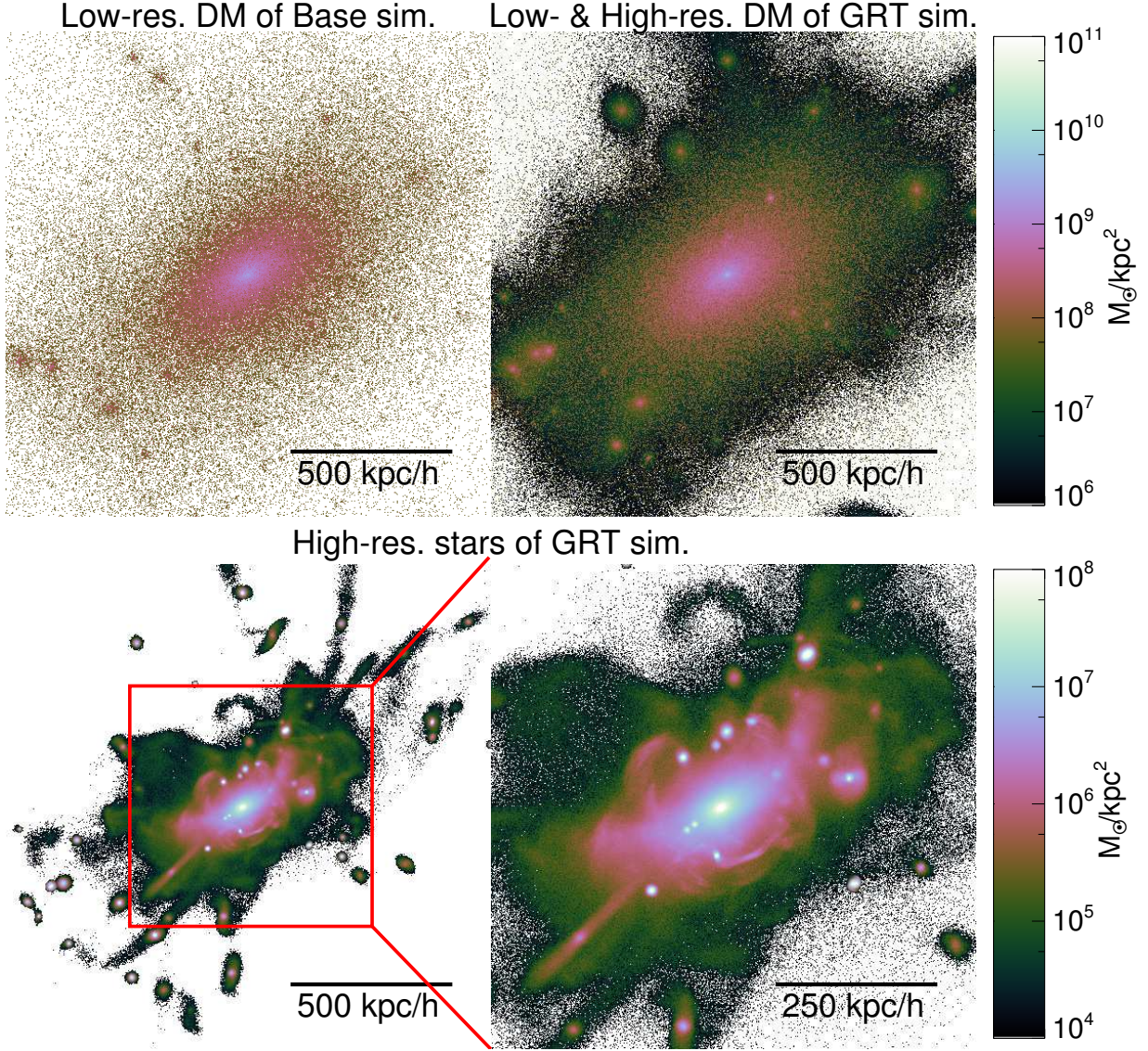}
\caption{The upper left and right panels show DM structures inside the virial radius ($R_{vir}$) of the GRT cluster at $z=0$ in the base simulation and the GRT simulation. The lower left and right panels show the stellar structures in $R_{vir}$ and 0.5$R_{vir}$ of the GRT cluster. The structure of each panel is colored by DM and stellar surface mass density. The color scale for DM and stellar structures is shown next to the right panels.}
\label{fig:structure}
\end{figure*}

In order to complete the GRT simulation, we perform our multi-resolution re-simulation until $z=0$ using the cosmological simulation code, Gadget-3 \citep{springel2005}. We choose the time interval of the snapshots to be the same as that of the low-resolution simulation.
The GRT simulation has a gravitational softening length for the high-resolution DM and star particles of $\sim$100~pc$/h$ and $\sim$10~pc$/h$, respectively. With such a high spatial resolution, we can well resolve the self-gravity of the disks radially and vertically out of the disk plane, and accurately follow the tidal stripping process. The gravitational softening length for the low-resolution DM is 2.3 kpc/h, which is the same as in the base simulation.
The high-resolution particle mass for DM and star is 5.4$\times$10$^{6}$~M$_{\odot}/h$ and 5.4$\times$10$^{4}$~M$_{\odot}/h$, respectively. The mass of high-resolution DM (or star) particles is $\sim$ 200 (or 20,000) times smaller than that of the low-resolution DM particles.

The mass of the high-resolution star particles is similar to the baryon mass or minimum mass of the star particles in Illustris TNG50 or NewHorizon simulations \citep{nelson2018,dubois2020}.
However, those two different simulations used 130 million and 40 million CPU hours to simulate until $z=0$ and $z=0.7$, respectively.
Note there is only one cluster in Illustris TNG50 and one group in NewHorizon, so it is impossible to consider the influence of cluster-to-cluster variations on the diffuse light using these simulations.
On the other hand, our GRT simulation used only 50,000 CPU hours to trace the evolution of one cluster until $z=0$.
Thus, the GRT opens up the possibility of conducting large statistical studies of tidal stripping and ICL formation with a  live distribution of star particles in groups and clusters for the first time.

Figure \ref{fig:structure} shows DM or stellar structures of the GRT cluster, colored by surface mass density.
Clearly, the surface density where we can make reliable predictions about the DM structures is much lower ($<$10$^{8}$~M$_{\odot}/$kpc$^2$) in the GRT simulation compared to in the base simulation (compare the upper left and right panels).
The stellar structures of the GRT cluster in the lower left panel show the diffuse light and tidally disrupted structures surrounding the BCG in the center.
In the GRT simulation, we do not insert a high-resolution galaxy model for the BCG. Instead, the BCG can build up by merging many high-resolution galaxies because we start the re-simulation early when the cluster is a tiny fraction ($\sim$ 0.1\%)  of its z=0 mass. The lower right panel shows the complex structures in the central region with a zoom-in view for more detail.

\begin{figure}
\centering
\includegraphics[width=0.45\textwidth]{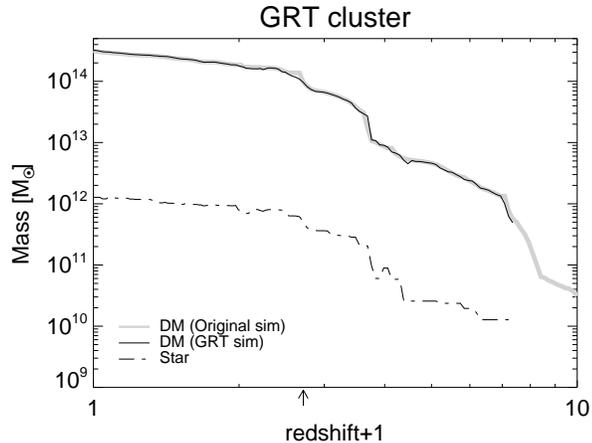}
\caption{Mass growth history of the GRT cluster. The thick gray and thin black solid lines represent the growth history of the cluster in the base simulation and the GRT simulation respectively. The dash-dotted line indicates the mass growth of the stars in the GRT cluster. The black arrow below the x-axis means the epoch of the last major merger of the cluster.}
\label{fig:mass_growth}
\end{figure}

Overall, we find the mass evolution of the GRT cluster in the GRT simulation (the thin solid line) follows the evolution of the base simulation (the thick solid line) well (Figure \ref{fig:mass_growth}).
The GRT cluster undergoes the last major merger at $z\sim1.7$ (the arrow below the x-axis) and accretes half of its final mass by $z\sim1.4$.
As the stars are supplied by the high-res. galaxies that fall into the cluster, the stellar mass growth (the dash-dotted line) shows similar growth episodes as the DM mass growth.

In summary, the GRT focuses on accurately modeling the tidal stripping of galaxies in the tidal field of the cluster and its sub-structures within a full cosmological context. As the GRT is a multi-resolution cosmological N-body re-simulation, this allows us to place the high resolution where it is most needed. 
Typical large volume hydrodynamical cosmological simulations (e.g., Illustris TNG100, EAGLE, etc.) have a spatial resolution of $\sim$1~kpc. Therefore, the small disks ($\sim$1~kpc) of low mass galaxies would be almost completely unresolved. In practice, \citet{genel2018} showed that their model galaxies (M$_{star}>$ 10$^{9}$~M$_{\odot}$) tend to have larger disks compared to observed galaxies of the same mass. During tidal mass loss, the stellar disks begin to be stripped much more easily for larger, more extended disks \citep{smith2016}, and eventually this will affect the growth and evolution of more massive galaxies that hierarchically form from them. With our high spatial resolution (10~pc$/h$), and precise control over the size of the stellar disks that we insert, our galaxies are more realistically robust to external tides. This issue is discussed further in Section \ref{sec:heating}.
Moreover, the low mass of star particles (5.4$\times$10$^{4}$~M$_{\odot}/h$) means we can make accurate predictions about very low surface brightness features. Indeed, this mass resolution allows us to model features that are approximately three mag arcsec$^{-2}$ lower surface brightness than in Illustris TNG100. The use of abundance matching also ensures our galaxy masses match the observed galaxy masses over a broad range of redshift. The low computational cost enables us to model ICL evolution and tidal features (tidal tail, tidal stream, shell-like structure, etc.) with live star particles for large samples of clusters. This means we can study dependencies on cluster variations, such as merger history, for the first time with a statistically significant sample of clusters.

\section{Testing the GRT or verification of the GRT}
\label{sec:test}

In this section, we run some tests to test whether the GRT and some of the assumptions in the technique are appropriate to investigate the formation of the ICL.
In Section \ref{sec:am}, we investigate the consequences of our choice of halo abundance matching when replacing galaxies for matching the total stellar mass observed in groups and clusters over the redshift range from $z=1$ to $z=0$.
In Section \ref{sec:heating}, we test if the mixing of high-resolution particles with low-resolution particles that can occur in the GRT simulation is enough to cause significant artificial disk heating.

\subsection{Abundance matching}
\label{sec:am}

As described in Section \ref{sec:GRT}, we insert a full high-resolution model galaxy, with halo and stellar disk component, just before a halo begins to lose mass in the tidal potential of another more massive halo (for example, a group or cluster halo).
We decide the stellar mass of the galaxy at this time following a stellar-to-DM halo mass relation derived from halo abundance matching. In general, we use the time-evolving halo abundance matching model of \citet{behroozi2013c} for galaxies over a wide range of redshift up to $z=8$.

\begin{figure*}
\centering
\includegraphics[width=0.9\textwidth]{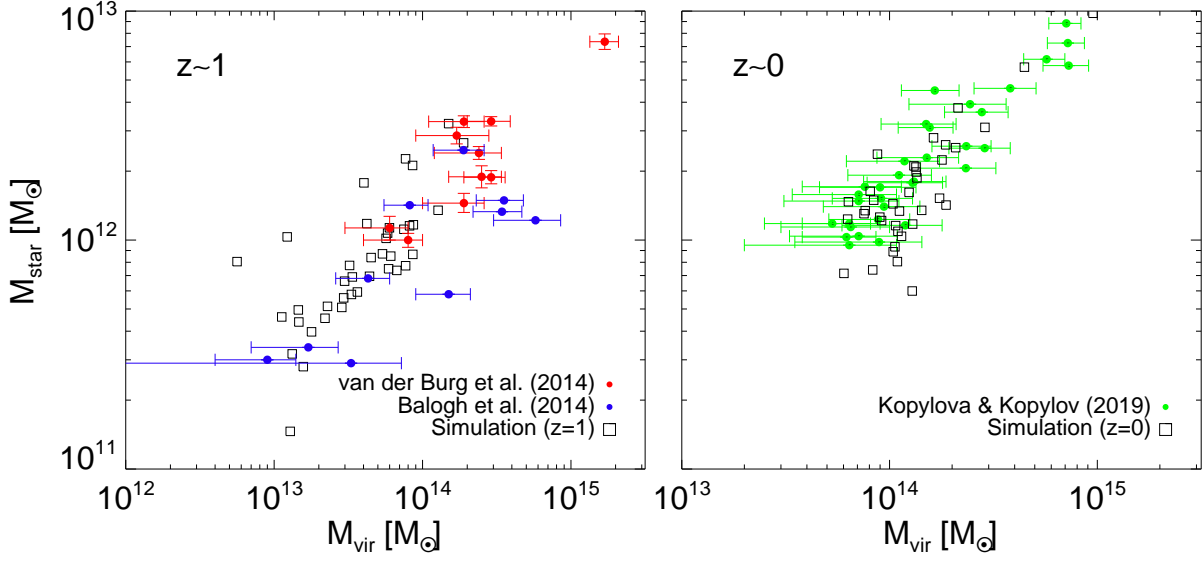}
\caption{The total stellar mass contained within the halo of simulated groups/clusters (black open squares) versus the observed groups/clusters (\citet{burg2014}, \citet{balogh2014}, \citet{kopylova2019} indicated by red, blue, and green filled circles respectively)  at $z\sim1$ (the left panel) and $z\sim0$ (the right panel).}
\label{fig:ms}
\end{figure*}

To investigate whether this choice of halo abundance matching model provides satisfactory results in the GRT simulation, we compare the total stellar mass of the 44 clusters of M$_{vir}(z=0)>$10$^{14}$~M$_{\odot}/h$ within one cosmological box of the base simulation with the observed total stellar mass of groups and clusters at $z\sim0$ and $z\sim1$ \citep{balogh2014,burg2014,kopylova2019} (Figure \ref{fig:ms}).
Hereafter, we refer to our modeled 44 clusters as `sample clusters'.
The total stellar mass of sample clusters is measured as the sum of the stellar mass of all high-res galaxies falling into the clusters until $z=0$ or $z=1$.
\citet{balogh2014} and \cite{burg2014} studied eleven massive groups and clusters (the filled blue circles) and ten more massive clusters (the filled red circles) at $z\sim1$ using the GEEC2 survey and GCLASS survey.
The stellar mass of the 34 observed groups and clusters at $z\sim0$ (the filled green circles) are determined by the 2MASX catalog \citep{kopylova2019}.
The black open square symbols indicate the sample clusters at $z=1$ and $z=0$ (the left and the right panel, respectively).
These observational studies show that the total stellar mass is a good proxy for the virial mass of the groups and clusters.
In Figure \ref{fig:ms}, we find that the stellar mass-halo mass relation of the sample clusters provides a good match to the observations at both redshifts. 
Although we do not show the result, we also find that the GRT provides a good match of the stellar mass – halo mass relation of the observed groups \citep{connelly2012}.

While we clearly get good results with the \citet{behroozi2013c} abundance matching model, we note that there can be some variation, at fixed halo mass, and across redshift, between the various abundance matching models \citep{moster2013,behroozi2013c,rodriguez-puebla2017,girelli2020}. Indeed, we find that other abundance matching models \citep{moster2013,rodriguez-puebla2017,girelli2020} estimate up to 30\% less or 20\% more total stellar mass of the sample clusters than the abundance matching model of \citet{behroozi2013c}. 
However, despite the differences in the total stellar mass, we find that the fraction of the stellar mass of disrupted halos (which contributes to the ICL) among the total stellar mass in the cluster differs by only 2\% among the different abundance matching models. Therefore, while the total stellar mass might vary, we do not expect the choice of the abundance matching model will significantly affect the ICL fraction. Given that the total stellar masses were best matched with the abundance matching model of \citet{behroozi2013c}, we conclude that this is an appropriate choice for the GRT simulation. 

\subsection{The stability of disk galaxy models}
\label{sec:heating}

\begin{figure*}
\centering
\includegraphics[width=0.95\textwidth]{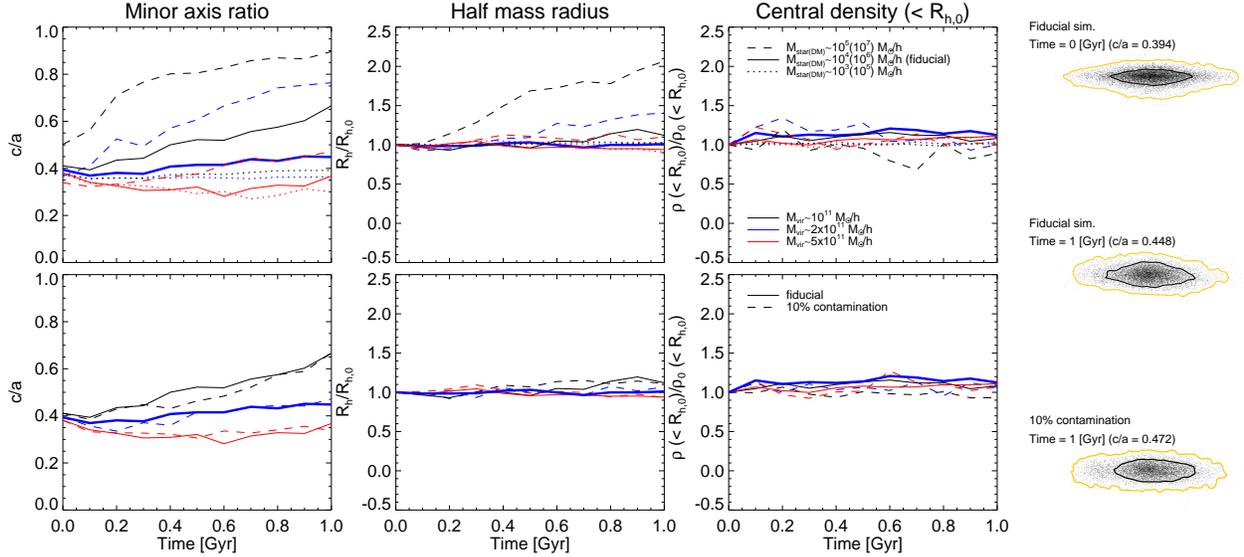}
\caption{Time evolution of the isolated disk galaxy. The left three columns show the evolution of the minor axis ratio (c/a), half mass radius ($R_h$), and central density, $\rho (<R_{h,0})$. $R_h$ and $\rho (<R_{h,0})$ are normalized by their initial value, $R_{h,0}$ and $\rho_0 (<R_{h,0})$.} The black, blue, and red lines represent the halos of M$_{vir}\sim$10$^{11}$, 2$\times$10$^{11}$, and 5$\times$10$^{11}$~M$_{\odot}/h$, respectively. In the three left upper panels, dashed, solid, and dotted lines indicate the simulations with the lower, fiducial, and higher mass resolution. The mass of the star and DM particles in each simulation is shown the top of the third panel. In the bottom panels, solid and dashed lines indicate the fiducial models and the halos that have 10\% of their inner halos ($R<10$~kpc$/h$) in the form of low-resolution DM particles. The thick solid lines indicate the most common case in the GRT simulation. The right three rows show the initial structure of the disk in the most common halo and disk structures at 1~Gyr with or without contamination. The yellow and black isodensity contours indicate the outskirts and $R_h$ of the disk. The minor axis ratio, c/a, of each disk is shown on the top of each figure.
\label{fig:dice}
\end{figure*}

It is inevitable that some level of artificial heating will arise in numerical simulations of galaxies. Two-body interactions from the difference in mass between the disk and DM particles can result in artificial morphological transformation, or even scattering of star particles out of the disk in extreme cases \citep{velazquez2005,fujii2011,sellwood2013}.
As we wish to study the formation of the ICL in the GRT simulation, we should pay particular attention to any artificial increase in the size of the modeled disk galaxies. This is because larger disks are more susceptible to tidal stripping \citep{smith2016} which then feed the growth of the ICL.
We also pay special attention to the issue because the GRT places high-resolution particles in a background of low-resolution particles. Therefore we wish to understand the consequences well.

By inserting the high-res. galaxy when it has M$_{peak}$, we try to significantly reduce the amount of low-resolution particles that contribute to the growth of the halos (i.e., there should be little additional accretion once the halo stops growing in mass).
In practice, we find this strategy is effective in the GRT simulation. 80\% of the high-res. galaxies do not have any low-resolution DM particles bound to their halos during their lifetime.
Even though the others do have some bound low-resolution DM particles, we measure that less than 5\% of their inner halos ($R<10$~kpc$/h$) are in the form of bound low-resolution particles by mass.

To test the influence of this level of contamination and to better evaluate numerical heating, we perform simulations of an isolated disk galaxy with three different masses (M$_{vir}=$10$^{11}$, 2$\times$10$^{11}$, and 5$\times$10$^{11}$~M$_{\odot}/h$).
As discussed in Section \ref{sec:GRT}, the model with M$_{vir}=$10$^{11}$~M$_{\odot}/h$ is the lowest mass galaxy that is used for low-resolution halo replacement in the GRT simulation. They are built in the same way as in the GRT, for a $z=0$ galaxy.
For each model, we perform four different simulations during 1~Gyr, with the three different mass resolutions and the contamination by low-resolution DM particles.
Given the disk models are isolated, we monitor the evolution of the minor axis ratio (c/a), half mass radius ($R_h$), and central density, $\rho (<R_{h,0})$, to look for possible indications of artificial disk heating. We use half mass radius and central density normalized by their initial value, $R_{h,0}$ and $\rho_0(<R_{h,0})$.
The minor axis ratio, c/a, of each disk is calculated using the eigenvalues of the disk reduced inertia tensor limited to within 10~kpc$/h$. We iterate this calculation until the axis ratio converges (see \cite{vera-ciro2011}).The central density is measured using the total mass (DM+star) within the initial stellar half mass radius ($R_{h,0}$). The central density is important for determining the robustness to tides. If the central density of the disk gradually decreases, then tides will be more disruptive. Furthermore, as the previous studies \citep{penarrubia2008,smith2016} have shown that the more radially extended a galaxy is within a halo, the easier it is to tidally strip the disk. Therefore, an artificial increase in $R_h$ could cause artificially increased stripping of stars from the outer disk. Because we generate the initial conditions in the equilibrium state, these changes are related to artificial heating.

The left six panels of Figure \ref{fig:dice} show the evolution of c/a, $R_h$, and $\rho (<R_{h,0})$ of the three different galaxy models. The upper panels show how artificial heating is affected by the mass resolution of the DM and star particles.
The black, thick blue, red solid lines indicate the halos having M$_{vir}=$10$^{11}$, 2$\times$10$^{11}$, and 5$\times$10$^{11}$~M$_{\odot}/h$, respectively. The mass and spatial resolutions of DM and star particles are the same as that of the high-resolution particles in the GRT simulation. We call these the fiducial models. The thick blue lines indicate the most common mass halos in the GRT simulation. To investigate the effects of the mass and spatial resolutions on the disk evolution, we also perform the simulations with the lower and higher number of particles by a factor of 10 about the value we use in the fiducial models (the dashed and dotted lines), while keeping the star-to-DM particle mass ratio constant. The different color lines indicate halos with the same mass as the fiducial models.

The results show that the central density of all halos is maintained. It indicates that the DM halo is not affected by artificial heating even in the low-resolution halos. In the case of the stellar disk, as might be expected, there is more artificial heating when there are less particles. However, at our fiducial resolution, the evolution over 1~Gyr in disk radius is negligible, and the evolution in disk thickness is minor, except in the lowest mass model case. Fortunately, disk thickness has been shown to have a negligible impact on the efficiency of stellar stripping \citep{smith2021}. More importantly, the fact that $R_h$ and $\rho (<R_{h,0})$ are constant is very reassuring as it means that any artificial heating that exists will not artificially enhance the amount of stellar stripping occurring in the GRT simulation. Furthermore, these low mass galaxies contribute less than $\sim$8\% of the total stellar mass of all the high-res. galaxies in 44 sample clusters. 
Even though we find that most of the low-resolution particles are not bound and pass through the galaxy in most high-res. galaxies, in 20\% of high-res. galaxies, a few bound low-resolution particles are found in their inner halos. Therefore, we pay attention to whether the stellar disk of these contaminated halos in significantly affected by artificial heating due to the low-resolution particles.

To investigate how this contamination affects the evolution of disk galaxy, we perform a set of controlled simulations, including contamination by low-resolution DM particles in the central 10~kpc/h of the halos (the three left bottom panels of Figure \ref{fig:dice}). We replace 10\% of the high resolution DM particles with low-resolution particles out to $R_{vir}$. The choice of 10\% should be considered an extreme case in the GRT simulation, as even the most contaminated high-res. galaxy has only 5\% of its central 10~kpc/h in the form of bounded low-resolution DM particles. 

The solid and dashed lines in the lower panels of Figure \ref{fig:dice} indicate the fiducial models (no contamination) in the upper panels and the 10\% contaminated models, respectively. The lower left panel shows that the evolution of the minor-to-major axis ratio in the contaminated halos is similar to the fiducial models.
The lower right panels show that, once again, $R_h$ and $\rho (<R_{h,0})$ are constant, despite the contamination.
Overall, this indicates that even with the quite extreme 10\% contamination, we get little artificial heating.

The right column of Figure \ref{fig:dice} shows the edge-on disk of the most common halo initially (top), after 1~Gyr evolution (middle), and with 10\% contamination (bottom). Disk thickening from the low resolution contamination is quite mild, and there is no indication of the disk size evolving.

In summary, using the GRT, we can combine high mass resolution (5.4$\times$10$^{4}$~M$_{\odot}/h$) and high spatial resolution (10~pc$/h$), and we find that the level of the artificial heating present is not significant for our studies of the ICL formation.
As we will discuss in Section \ref{sec:discussion}, this is a significant improvement on the  GRT-like implementation in \citet{rudick2006} which must have resulted in the accretion of more low-resolution particles because of the fixed epoch at which the galaxies were replaced. 

\section{ICL detection methods}
\label{sec:ICL}

Since there is no unique definition for the ICL, previous studies for the ICL have used various detection methods \citep{feldmeier2004,gonzalez2007,rudick2011,burke2015,longobardi2015,canas2020}.
In this section, we introduce the three main detection methods and use them to measure the fraction of cluster light in the ICL and in the BCG \citep{mihos2005,rudick2011,burke2013,contini2014,contini2018}.
Tidal streams generally satisfy the criteria for ICL stars in all three detection methods.
The surface brightess limit method is described in Section \ref{sec:SBL}, and Section \ref{sec:rho} and Section \ref{sec:rockstar} present the 3D density and boundness limits methods, respectively.
Section \ref{sec:fraction} shows the time evolution of the ICL and BCG fraction of a single cluster using various detection methods.

\subsection{Surface brightness limit}
\label{sec:SBL}
Because the ICL is connected to the outer region of the galaxies, it can be challenging to separate the ICL from the galaxies observationally.
For this, many observational studies defined a surface brightness limit (SBL) below which the light is considered as ICL \citep{feldmeier2004,montes2014,montes2018,presotto2014,burke2015,ko2018}.
Although the SBL for the ICL has less physically meaningful, it is a convenient method for directly comparing the properties of the ICL in observations with those in simulations \citep{rudick2006,rudick2011,cui2014,tang2018}.
There is no definite SBL of the ICL, but we use a V-band surface brightness of $\mu_{V}=26.5$ mag arcsec$^{-2}$, which is consistent with the value that many different previous works have chosen \citep{rudick2006,rudick2011,cui2014,presotto2014,tang2018}.
In this scheme, the galaxies naturally are defined as the regions brighter than the SBL.
As the SBL links the diffuse light which fills the space between the galaxies, we can separate the galaxies into individual groups using a 2D friends-of-friends (FOF) algorithm.
The BCG is defined as the largest 2D FOF group among the galaxies in the central region ($r<R_{vir}$) of the cluster. The fraction of the ICL/BCG is the ratio of the total luminosity within the grids that satisfy the definition of the ICL /BCG to the total luminosity within $R_{vir}$ of the cluster.

To generate the surface brightness maps of the GRT cluster, we use the projected two-dimensional stellar particle distribution using all stellar particles in $R_{vir}$ of the cluster, without considering if the particles are bound to the cluster or not.
We use the stellar distribution projected onto the xy-plane. In practice, we find that the exact choice of projection plane only affects the ICL fraction by less than 2\%.
After projecting the particles onto the 2D plane, the stellar particles are binned into a grid of cells with length $D$=800~pc. We find that the ICL fraction is not sensitive to the choice of the length $D$ (increasing or decreasing $D$ by a factor of two alters the ICL fraction by less than 2\%).
Because we perform collisionless N-body simulations in this work, we cannot know the exact age or metallicity of each stellar particle without making many assumptions.
This means it is difficult to specify a mass-to-light ratio for each stellar particle.
Therefore, for simplicity, we use a constant V-band mass-to-light ratio of 5~M$_{\odot}/$L$_{\odot}$ to convert from the total stellar mass in each cell to the luminosity. This constant V-band mass-to-light ratio is a typical value for old stellar populations \citep{bruzual2003,longhetti2009,burke2012}.

Next, the surface brightness maps are convolved with a 2D Gaussian point spread function (PSF) of 51$\times$51 pixels, matching the PSF of the SDSS to mimic the observed galaxy clusters. 
To compare the time-evolution of the properties of the ICL and BCG with the other two different detection methods, we use the rest-frame surface brightness without the redshift dimming factor when we generate the surface brightness maps in this work.

\subsection{3D density limit}
\label{sec:rho}

We can obtain more accurate ICL/BCG fractions if we use the full 3D positions of the particles in the simulation, although we cannot directly compare these 3D measures with those of the observations easily.

\begin{figure}
\centering
\includegraphics[width=0.45\textwidth]{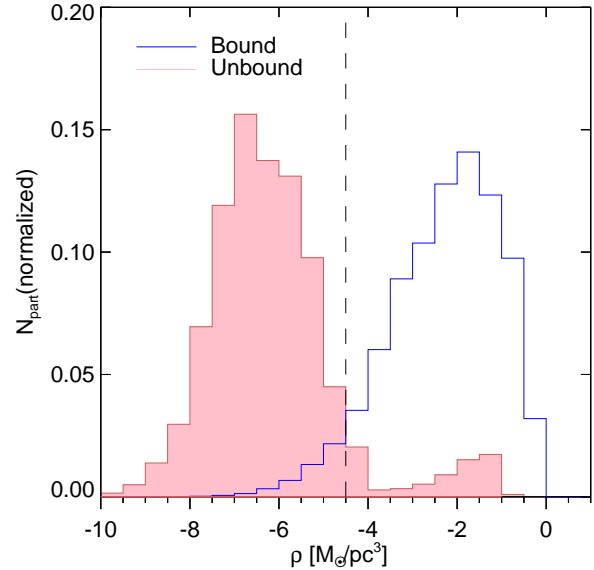}
\caption{Histogram of the bounded (blue) and unbounded (red) particles in the satellites' galaxies. Each histogram is normalized by the total number of the bounded and unbounded particles. The vertical dashed line indicates $\rho_{thresh}$.}
\label{fig:rho}
\end{figure}

Because the ICL is composed of a diffuse component, we expect it to have a low 3D density.
Therefore, the simplest method is to define a 3D density limit $\rho_{thresh}$ below which star particles are considered part of the ICL.
This density-based definition of the ICL was introduced by \citet{rudick2009}.
In \citet{rudick2009}, the density of a stellar particle was defined as the mass density of the stellar particles within the particle's 100th nearest neighbor.
To exclude stellar particles that are only temporarily separated from their host galaxy by the limit, a minimum consecutive time period, $t_{min}$, is adopted. \citet{rudick2009} chose $\rho_{thresh}$ and $t_{min}$ as $10^{-5}$~M$_{\odot}$~pc$^{-3}$ and 200~Myr, respectively.
In this work, we use the same method to measure the density and the same value of $t_{min}$, but we define $\rho_{thresh}$ based on the density distribution of the bounded and unbounded particles in the satellites' galaxies of the GRT cluster (Figure \ref{fig:rho}).
The blue and red histograms indicate the density distribution of the bounded and unbounded particles in the satellites' galaxies, normalized by their total numbers.
We choose $\rho_{thresh}$=$10^{-4.5}$~M$_{\odot}$~pc$^{-3}$ because this best separates the density distribution of the bounded and unbounded particles in the satellites.

In this detection method, the BCG is defined as the largest 3D FOF group, which is denser than $\rho_{thresh}$ and in the central region of the GRT cluster.
This definition for the BCG is similar to that of the SBL method but using the 3D FOF algorithm instead of a 2D FOF algorithm.
Finally, the ICL/BCG fraction is the ratio of the number of star particles satisfying the definition of ICL or BCG divided by the total number of star particles within $R_{vir}$ of the cluster.

\subsection{Boundness limit using 6DFOF algorithm}
\label{sec:rockstar}

The last method to identify the ICL and BCG is using the six-dimensional phase-space halo finder, ROCKSTAR \citep{behroozi2013a}.
The most theoretical definition of the ICL is the stellar components that are gravitationally bound to the cluster but not bound to any one galaxy.
This means that the ICL is the component that can be kinematically too hot or physically distinct to be part of a galaxy. Thus the ICL and galaxies are separable in the phase-space plane.
Because the BCG is also composed of star particles that are only bound to the DM halo of the cluster, ROCKSTAR can separate the BCG from the galaxies in the cluster's sub-halos.

ROCKSTAR first finds over-dense regions using the 3D FOF algorithm, and particles within the linking length of each particle, $b=0.28$, are attached to the same 3DFOF group. 
For each 3DFOF group, the hierarchical FOF subgroups are repeatedly built in the 6D phase-space plane using a phase-space linking length, $f=0.7$.
The values which we use are the recommended values in \citet{behroozi2013a}.
They suggest that these values are suitable for removing the spurious subgroups and finding all particles near $R_{vir}$ of halos.
A seed halo is generated in the deepest level of the hierarchical FOF subgroups, and particles are assigned to the closest seed halo in the phase-space plane. 
In the same 3D FOF group, a smaller halo is defined as a sub-halo of the closest larger halo.
This allows ROCKSTAR to establish the relation between the cluster and its sub-halos.
Among the particles assigned to the cluster and sub-halos, we identify the particles only bound to the halo of the cluster to define the combined ICL and BCG.
Therefore, the fraction of the combined ICL and BCG is the ratio of the number of stellar particles which are only bound to the cluster compared to the total number of stellar particles within $R_{vir}$ of the cluster. 

\subsection{The fraction of the ICL and BCG}
\label{sec:fraction}

A number of studies have looked at the evolution of the ICL and BCG fraction and tried to correlate it with the dynamical state or mass of the clusters \citep{conroy2007,burke2015,contini2014,contini2018}.
However, each detection method introduced earlier could show different results for the ICL and BCG fraction if different methods are used to define them.

\begin{figure*}
\centering
\includegraphics[width=0.9\textwidth]{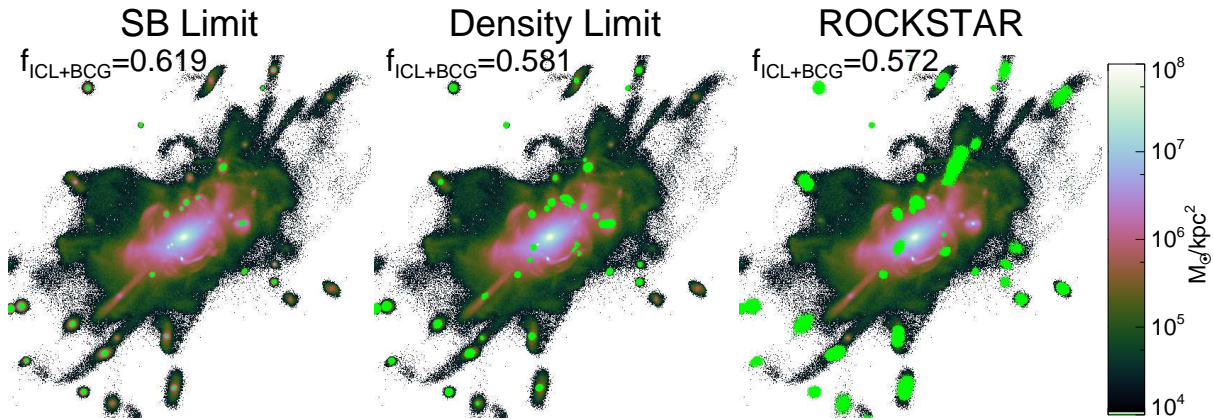}
\caption{Stellar structures of the GRT cluster using the surface brightness, 3D density, and boundness limit methods (From left to right panel). The structures of the BCG and ICL defined using the different detection methods is colored by stellar surface mass density. The lime regions indicate the satellites defined by each method. The color scale of the structures is the same as that of Figure \ref{fig:structure}. The fraction of the BCG and ICL is shown at the top of each panel.}
\label{fig:icl_map}
\end{figure*}

Figure \ref{fig:icl_map} illustrate the stellar structures ($<0.5R_{vir}$) of the GRT cluster using the different detection methods. The color map indicates the stellar surface density of the inner R$<0.5R_{vir}$ of the cluster (and is identical in all panels). However, the fraction of the total stellar mass contained within $R_{vir}$ that is assigned to be ICL+BCG (as opposed to satellites) differs between the panels (see the fraction f$_{\rm{ICL+BCG}}$ indicated in upper left of each panel).

The lime color overlaid highlights pixels containing satellite star particles as identified by each method. Comparing between the panels, it is clear that the `SB limit' method seems to miss some small satellite members, perhaps because of projection effects, that the `Density Limit' method recovers. The binding criteria of `ROCKSTAR' method allows it to recover bound stars in tidal streams that emanate from satellite members. Furthermore, we can see that this method classifies the particles in regions that are less dense than $\rho_{thresh}$ as the satellite members because of the binding criteria.

\begin{figure}
\centering
\includegraphics[width=0.45\textwidth]{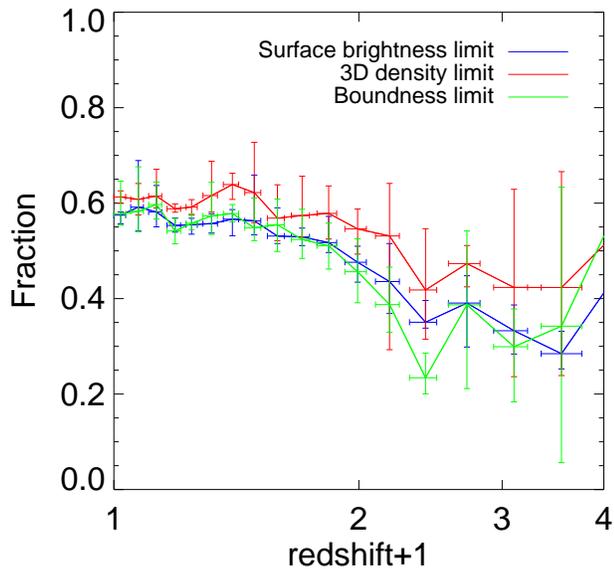}
\caption{Evolution of the ICL and BCG fraction, f$_{\rm{ICL+BCG}}$. The figure shows the evolution of f$_{\rm{ICL+BCG}}$ measured by the surface brightness (the blue solid line), the 3D density (the red solid line), and boundness (the green solid line)} limit methods. The x-axis and y-axis error bars indicate the range of each time bin and the ICL and BCG fraction.
\label{fig:ficl}
\end{figure}

Figure \ref{fig:ficl} shows that the differences in recovered f$_{\rm{ICL+BCG}}$ began early and have continued since high redshift until now.
Although there is an offset among the three different detection methods, overall, the evolution of f$_{\rm{ICL+BCG}}$ shows similar features.
We can see that f$_{\rm{ICL+BCG}}$ is increasing steadily from $z\sim3$ even if there is some noise prior to $z=1$, in all of the detection methods.
However, the fractions remain roughly constant after $z\sim0.8$. 
We find these fractions are similar to the maximum fraction ($\sim$ 60\%) of the stellar mass within the BCG and ICL of observed clusters (M$_{200c}>$ 10$^{14}$~M$_{\odot}$) at $z\sim0.1$ \citep{gonzalez2013,kravtsov2018}. We convert the observational data from M$_{500c}$ to M$_{200c}$ to compare with the mass of the GRT cluster. Here, M$_{500c}$ and M$_{200c}$ are defined as the mass within the radius where the mean mass density is 500 and 200 times the critical density, respectively.
The similarity of the time evolution in the three detection methods indicates that they respond similarly to events such as the merger, disruption of the satellites, etc.
Nonetheless, due to the offset of f$_{\rm{ICL+BCG}}$ between the detection methods, we should be careful when comparing studies using the different methods.

In a follow-up study paper, we will perform a more detailed study, considering the various formation channels and evolutionary trends of the ICL, in a larger sample of clusters with various masses and dynamical states.
Moreover, we will investigate if the offsets in f$_{\rm{ICL+BCG}}$ between the detection methods are universal in all clusters or depend on their differing growth histories.

\section{Discussion}
\label{sec:discussion}

\subsection{Comparison with previous work}
\label{sec:rudick}

A different version of the galaxy replacement technique (GRT) was first used to study the ICL by \citet{rudick2006}. However, we have made significant adaptations and improvements to the original approach that we now describe here.

(1) The time to insert the galaxy - \citet{rudick2006} performed the cosmological N-body simulations for the three different clusters.
For each cluster, they identified the halos at $z=2$ that will fall into the clusters until $z=0$ and inserted the higher resolution galaxy models (stars and DM halos) in place of the low-resolution halos all at a fixed time ($z=2$).
In contrast, we replace the galaxies when the halos reach M$_{peak}$, which can vary in time between $z=0.005$ to 7.9 for our sample clusters.
As we discussed in Section \ref{sec:heating}, this choice heavily reduces the accretion of low-resolution DM particles which could potentially enhance artificial disk heating in \citet{rudick2006}.

(2) The galaxy model - because \citet{rudick2006} inserted all galaxies at $z=2$, some halos can have satellite galaxies as well as the central galaxy. 
To reproduce this, they decided the number of galaxies in the halos using the halo occupancy (HOD) model based on M$_{vir}$ of the halos. 
The stellar mass of the halos was fixed at a tenth of the DM halo regardless of the mass of the halos.
If there were multiple galaxies in the halos, the mass of the individual galaxy was determined by a power-law mass function.
The morphology of the inserted galaxies was decided to be one of the elliptical and disk galaxy based on the local galaxy density.

As we mentioned in Section \ref{sec:GRT}, we insert one galaxy for each high-res. galaxy.
The stellar mass of each galaxy is determined by the abundance matching model of \citet{behroozi2013c}, meaning the galaxy masses are matched to the observations. Our choices result in the total stellar mass in the sample clusters matching observed groups and clusters at $z=0$ and $z=1$. 
Each inserted galaxy is a pure exponential disk galaxy, but as the galaxies begin to merge and interact with other galaxies, the morphological transformation will naturally occur in the hierarchical merging context, and galaxies will naturally fall into more massive hosts and become satellites.

(3) the starting point of the re-simulation - Because \citet{rudick2006} replaced the higher resolution galaxies with the original DM halos at $z=2$, the re-simulation only begins at $z=2$. The starting point of the GRT is typically between $z=6.5$ and 7.9 for the sample clusters. This means that the GRT can be used to follow the early build-up of the diffused light since high redshift.
Indeed, we find that the amount of the ICL is already above 10\% of the total stellar mass in the GRT cluster at $z=2$, the starting point of \citet{rudick2006}.
Therefore, the GRT is useful for comparison with the observations beyond $z=1$ \citep{ko2018}.

\subsection{Future work with the GRT}
\label{sec:future_work}

\begin{figure*}
\centering
\includegraphics[width=0.9\textwidth]{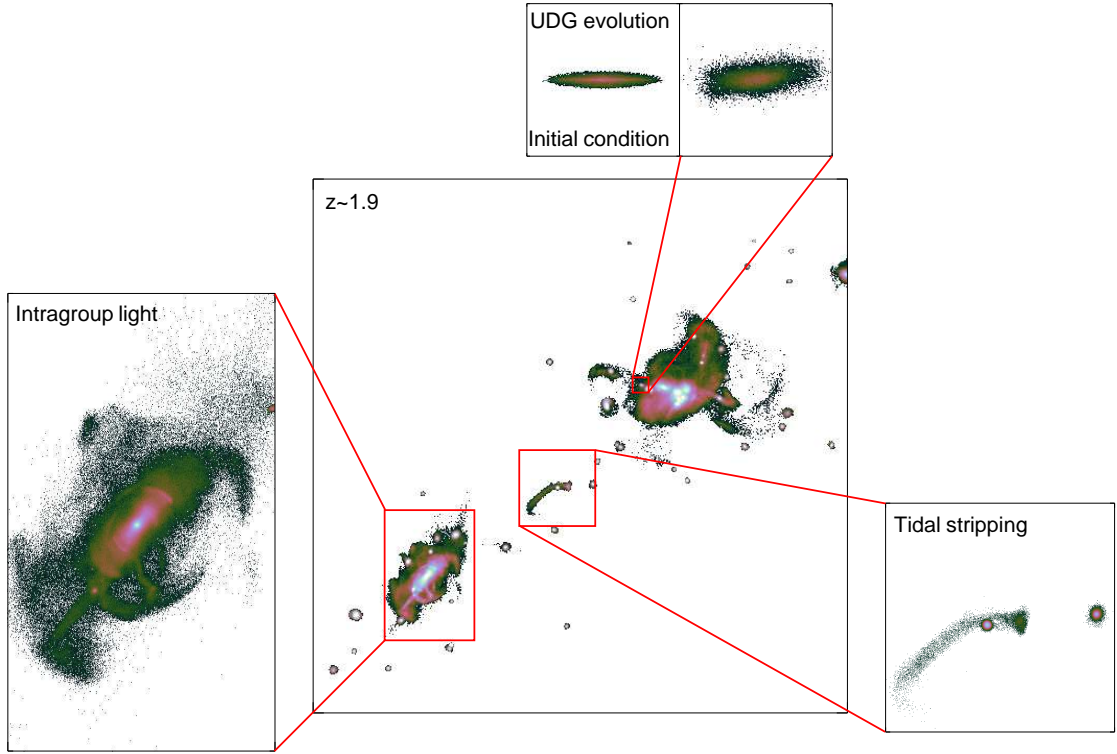}
\caption{Structures of the GRT cluster at the epoch of the last major merger. In the central panel, the GRT protocluster is in the upper right and a group-sized halo companion is in the bottom left. The left panel indicates the brightest group galaxy and the intragroup light and tidally stripped structure is in the right panel. The upper panel indicates a UDG candidate galaxy near the center of the GRT protocluster.}
\label{fig:future_work}
\end{figure*}

In this section, we discuss additional uses for the GRT that we will consider in the future.
In this paper, we have primarily focused on the evolution of the fraction of the ICL and BCG.
However, the GRT can trace the response of galaxies with a wide range of masses to gravitational tides in a fully cosmological context and since high redshift. 

Figure \ref{fig:future_work} shows various structures in the far outskirts of a cluster at the epoch of the last major merger of the GRT cluster.
The GRT protocluster is in the upper right of the main panel and a group-sized companion halo is shown in the bottom-left. The figure illustrates some alternative topics that the GRT could be applied to:

(1) The evolution of the galaxies at galaxy-group scales - in a hierarchical cosmological model, the cluster is built up by merging of smaller halos, which in turn contain their own satellites.
This means that these galaxies may have been pre-processed in the smaller group-sized halo before falling into the cluster. On the left panel, we can see the brightest group galaxy (BGG) and diffuse intragroup light (IGL) in a group-mass halo, even as early as $z=1.9$.

As the GRT inserts high-res. galaxies at their peak mass, many will be in the process of falling into groups \citep{mcgee2009,delucia2012,han2018}. The GRT will then accurately follow the tidal mass loss process inside the group scale halos with its high spatial and mass resolution.
Thus it is ideal for investigating the pre-processing of cluster galaxies from high redshift as well as the evolution of BGG and IGL and their contribution to the ICL as well.

(2) Tidal structures in clusters - In the strong tidal field of the cluster, satellites are disrupted and leave a variety of tidal features, including tidal streams (the bottom right panel), shell-like structures, and tidal tails.
Previous studies have shown their different features depend on the details of their orbits within the cluster, as well as being dependent on the type of the progenitors \citep{mihos2005,rudick2010,janowiecki2010}.
Similar dependencies also have been shown in previous studies for Milky Way-sized galaxies \citep{bullock2005,cooper2010,mancillas2019}.
The GRT traces the progenitor of these tidal structures from their formation time in a fully cosmological context. Our high spatial resolution allows us to accurately model the tidal stripping process. Moreover, the high mass-resolution of the GRT allows us to provide predictions on very faint structures (e.g. three mag arcsec$^{-2}$ deeper surface brightness limit compared to Illustris TNG100). Therefore, we can quickly and efficiently investigate the formation channels and properties of various LSB tidal structures, and in a variety of different environments, with varying merger histories.

(3) Ultra-diffuse galaxies - the ultra-diffuse galaxies (UDGs) have very low surface brightness, but their effective radius is larger than the dwarf galaxies of similar luminosity \citep{dokkum2015}.
Many previous studies have investigated the origin of the UDGs found in the cluster environment \citep{yozin2015,alabi2018,tremmel2020} and their survivability to cluster environmental mechanisms \citep{yozin2015,burg2016,mancerapina2018}.
We could use the GRT to study the survivability and evolution of the UDGs (the upper panel) to gravitational tides in a cosmological context.

A simulation using the GRT is a controlled simulation in a cosmological context.
Therefore, we are free to customize the mass or spatial resolution, choice of galaxies to replace, volume to consider, time period to model, galaxy models inserted, and various other settings for the simulation to improve the treatment of the scientific problem that we are interested in.
One example of this is for the UDG project above, where we could generate UDG model galaxies based on observations of UDGs. Given that some are quite low mass galaxies, we could choose to customize the base simulation for higher resolution and enhance the particle number in the high-resolution model, and carefully choose the location of where to insert them to investigate under which circumstances they can survive in a cosmological context. 

\section{Summary}
\label{sec:summary}

In this work, we introduce the Galaxy Replacement Technique (GRT). This is a new technique where we insert high-resolution models of galaxies into N-body cosmological simulations in order to study the impact of tides in a fully cosmological context. 

The GRT does not include computationally expensive hydrodynamical recipes. Instead, we replace low-resolution dark matter halos in the base cosmological simulation with high-resolution models of galaxies, including a dark matter halo and stellar disk. The properties of the halos are matched to the base simulation, except with greatly enhanced mass resolution. The mass of the stellar disk, by construct, will provide an excellent match to the observations of halo abundance matching at that redshift. The low mass of star particles means we can make predictions on very low surface brightness features ($\sim$three magnitudes below those in TNG100). The spatial resolution is very high (10~pc$/h$), much higher than can be found even in the latest cutting-edge hydrodynamical cosmological simulations (e.g., Illustris TNG50, NewHorizon), meaning we can accurately model the tidal stripping process. Numerically, the method is extremely fast. Modeling clusters in hydrodynamical cosmological simulations is notoriously computationally expensive, as they are such dense regions in a cosmological box. As such, TNG50 has only one cluster in its box, and NewHorizon has only one group. Based on the GRT we have modeled so far, we estimate we could model the full growth history of the ICL of thousands of clusters with the same computational time as provided to these simulations. This opens up the possibility of modeling ICL formation with a live star particle component in a statistically large sample of modeled clusters for the first time. This will enable us to study dependencies on various cluster properties (such as merger history) that are expected to be crucial factors for ICL formation. Therefore, the GRT will provide a powerful new tool for ICL and tidal mass loss studies.

However, we should be cautious of the limitations of this technique. In the GRT simulation, as we replace a low-resolution DM halo with a high-resolution galaxy when it has the peak mass, we would be unable to reproduce the formation of certain structures, for example such as a newly formed thin disk of stars that might form since the galaxy enters a more massive halo. In addition, the lowest mass galaxies (M$_{vir}\sim$10$^{11}$~M$_{\odot}/h$) in the GRT simulation show the significant disk thickening due to artificial heating in the current set-up for the ICL studies. Although these limitations are unlikely to be significant for the studies of the ICL formation, they might be more significant if we were focused on the evolution of the satellite galaxy properties. If we were interested in studying morphological transformation in low mass galaxies using the GRT, we would have to modify our high resolution galaxy models, such as forming them from higher resolution particles, as we discussed in Section \ref{sec:future_work}.

In this study, we present testing of the GRT technique and find that our choice of abundance matching recipe and choice of model resolution is well-suited for studying the ICL evolution.
In other words, we find our ICL fractions are not strongly dependent on the choice of abundance matching model that we use, and also we find that the high-res. galaxies do not suffer artificial tidal disruption due to their interaction with low-resolution DM particles.
As a demonstration of the technique, we present a GRT simulation of a single cluster of M$_{vir}\sim2\times$10$^{14}$~M$_{\odot}/h$ until $z=0$.
By $z=0$, we see the formation of spectacular low surface brightness features ($\mu_{V}<32$ mag arcsec$^{-2}$) such as streams, arcs, and shells surrounding the BCG. We then model the evolution of the ICL and BCG fraction, using the three different definitions for the ICL. In general, the ICL fraction increases steadily from $z\sim3$ and nearly remains constant after $z\sim0.8$. The final f$_{\rm{ICL+BCG}}$ is about 60\% at $z=0$. Comparing between the ICL definitions, we find that overall the time evolution of the fractions shows similar features at similar times, but clear offsets ($\sim5-20\%$) can be seen between the different ICL definitions.

In a follow-up study, we will expand on these preliminary ICL results and explore the various channels by which the ICL can form with a larger sample of clusters. We are also conducting a dedicated program to vastly expand the number of clusters we have modeled. Moreover, we are applying the GRT to various non-ICL related topics, as described in Section \ref{sec:discussion}. We hope the GRT will provide a valuable and powerful new tool to enhance our understanding of tidal mass loss and low surface brightness features in a fully cosmological context, from inside filaments of the large scale structure, to galaxy pairs, galaxy groups, clusters, or in the extreme case of cluster-cluster mergers.

\bibliography{main}{}
\bibliographystyle{aasjournal}

\end{document}